\documentclass{elsart}

\usepackage{graphicx}

\newcommand{\fig}[1]{~\ref{f:#1}}
\newcommand{\eq}[1]{~\ref{e:#1}}
\newcommand{\tab}[1]{~\ref{t:#1}}

\newcommand{\sech}{\,\mbox{sech}}
\newcommand{\csch}{\,\mbox{csch}}

\begin{document}

\begin{frontmatter}
\runtitle{New biased lattice random-walk model}   

\title{A new set of Monte Carlo moves for lattice random-walk models of biased diffusion}

\author[UofO]{Michel G. Gauthier}\ead{mgauthie@science.uottawa.ca},    
\author[UofO]{Gary W. Slater}\ead{gslater@science.uottawa.ca},          

\address[UofO]{University of Ottawa, 150 Louis-Pasteur, Ottawa, Ontario K1N 6N5, Canada}       


\begin{abstract}                          
We recently demonstrated that standard fixed-time lattice random-walk
models cannot be modified to properly represent biased diffusion
processes in more than two dimensions. The origin of this fundamental
limitation appears to be the fact that traditional Monte Carlo moves do
not allow for simultaneous jumps along each spatial direction. We thus
propose a new algorithm to transform biased diffusion problems into
lattice random walks such that we recover the proper dynamics for any
number of spatial dimensions and for arbitrary values of the external
field. Using a hypercubic lattice, we redefine the basic Monte Carlo
moves, including the transition probabilities and the corresponding
time durations, in order to allow for simultaneous jumps along all
Cartesian axes. We show that our new algorithm can be used both with
computer simulations and with exact numerical methods to obtain the
mean velocity and the diffusion coefficient of point-like particles in
any dimensions and in the presence of obstacles.
\end{abstract}

\end{frontmatter}

\section{Introduction} \label{s:intro}
Lattice Monte Carlo (LMC) computer simulations are often used to study
diffusion problems when it is not possible to solve the diffusion
equation. If the lattice mesh size is small enough, LMC simulations
provide results that are in principle arbitrarily close to the
numerical solution of the diffusion equation. In LMC simulations, a
particle is essentially making an unbiased random-walk on connected
lattice sites, and those moves that collide with obstacles are
rejected~\cite{Binder2,Barkema,Heermann,Binder}. The allowed Monte Carlo moves are usually displacements by
one lattice site along one of the $d$ spatial directions.

In the presence of an external field, one must bias the possible
lattice jumps in order to also reproduce the net velocity of the particle.
However, this is not as easy as it looks because one must also make
sure that the diffusion coefficient is correctly modelled along each of
the $d$ spatial directions. Using a Metropolis weighting factor~\cite{Binder2} does
not work because in the limit of large driving fields, all the jumps
along the field axis are in the same direction and hence the the
velocity saturates and the diffusion coefficient in this direction
vanishes. This approach is thus limited to weak fields, at best. A
better approach is to solve the local diffusion problem (i.e., inside
each lattice cell) using a first-passage problem (FPP)~\cite{Redner,Farkas1,VanKampen,Gardiner} approach, and to
use the corresponding probabilities and mean jumping times for the
coarser grained LMC moves. In this case, the mean jumping times are
shorter along the field axis, but one can easily renormalize the
jumping probabilities to use a single time step. In a recent
paper~\cite{PRL1}, we demonstrated that although this method does give
the correct drift velocity for arbitrary values of the driving field,
it fails to give the correct diffusion coefficient. The problem is due
to the often neglected fact that the variance of the jumping time
affects the diffusion process in the presence of a net drift~\cite{Bouchaud1}. LMC
models do not generally include these temporal fluctuations of the
jumping time, at least not in an explicit way. In the same article~\cite{PRL1}, we
showed how to modify a one-dimensional LMC algorithm with the addition
of a stochastic jumping time $\tau \pm \Delta\tau$, where the
appropriate value of the standard-deviation $\Delta\tau$ was again
obtained from the resolution of the local FPP. For simulations in
higher spatial dimensions $d>1$, it is possible to use our
one-dimensional algorithm with the proper method to alternate between
the dimensions as long as the Monte Carlo clock advances only when the
particle moves along the field direction~\cite{PRL1}.

LMC simulations of diffusion processes actually use stochastic methods to resolve
a discrete problem that can be written in terms of coupled linear equations.
Several years ago, we proposed a way to compute the exact solution of
the LMC simulations via matrix methods, thus bypassing the need for
actual simulations. This alternative method is valid only in the limit
of vanishingly weak driving fields, but it produces numerical results
with arbitrarily high precision. The crucial requirement of the method
is a set of LMC moves that have a common jumping time. Dorfman~\cite{Dorfman1,Dorfman2}
suggested a slightly different but still exact numerical method, and
the two agree perfectly at zero-field. More recently~\cite{Ogs9}, we extended our
numerical method to cases with driving fields of arbitrary magnitudes;
in order to do that, we used LMC moves that possess a single jumping
time for all spatial directions, but this forced us to neglect the
temporal fluctuations discussed above. As a consequence, our numerical
method generates exact velocities but fails to provide reliable
diffusion coefficients. Again, Dorfman's alternate method also give the
same velocities, but because the LMC moves do not include the proper
temporal fluctuations, neither method can be used to compute the
diffusion coefficient along the field axis. In summary, a fixed-time
LMC algorithm can be used with exact numerical methods to compute the
net velocity, but temporal fluctuations (and hence computer
simulations) must be used to compute the diffusion coefficient.

We recently solved the problem of defining a LMC algorithm with both a
fixed time step and the proper temporal fluctuations~\cite{PRL1}. This required the
addition of a probability to stay put on the current lattice site
during a given time step (of course, this change also implies a
renormalization of the jumping probabilities). This probability of
non-motion has a direct effect on the real time elapsed between two
displacements of the Brownian walker, and this effect can be adjusted
in order to reproduce the exact temporal fluctuations of the local FPP.
We showed that this new LMC algorithm can be used with Dorfman's exact
numerical method to compute the exact field-dependence of both the
velocity and the diffusion coefficient of a particle on a lattice in
the presence of obstacles. As far as we know, this is the first biased
lattice random-walk model that gives the right diffusion coefficient
for arbitrary values of the external field. Other models, such as the
repton model~\cite{Barkema}, are restricted to weak fields. Several
other articles (see, e.g.~\cite{Havlin,LopezSalvans,Bustingorry2}) report 
simulations of diffusive processes, but all of them appear to be limited to small biases.

Unfortunately, our LMC algorithm~\cite{PRL1} has a fatal flaw: for dimensions
$d>2$, some of the jumping probabilities turn out to be negative. This
failure suggests that there is a fundamental problem with this class of
models, or more precisely with standard LMC moves (however, note that
it is still possible to use computer simulations and fluctuating
jumping times $\tau \pm \Delta\tau$, as explained above). In other
words, it is impossible to get both the right velocity and the right
diffusion coefficient in all spatial directions (if $d>2$) when the LMC
jumps are made along a single axis at each step.

In this article, we examine an alternative to the standard LMC moves in
order to derive a valid LMC algorithm with a common time step for
spatial dimensions $d>2$. We suggest that a valid set of LMC moves
should respect the fact that motion along the different spatial
directions is actually simultaneous and not sequential. As we will
show, this resolves the problem and allows us to design a powerful new
LMC algorithm that can be used both with exact numerical methods and
stochastic computer simulations.

\section{The biased random-walk in one dimension}\label{s:1D}
As mentioned above, Metropolis-like algorithms are not reliable if one
wants to study diffusion via the dynamics of biased random-walkers on a
lattice~\cite{PRL1}. The discretization of such continuous diffusion
processes should be done by first solving the FPP of a particle between
two absorbing walls (the distance between these arbitrary walls is the
step size $l$ of the lattice). Indeed, completion of a LMC jump is
identical to the the first passage at a distance $l$ from the origin.
In one dimension, this FPP has an exact algebraic solution, and the
resulting transition probabilities (noted $\pm$ for parallel and
antiparallel to the external force $F$) are~\cite{Slater93}:
\begin{equation}
    \label{e:p1D}
    p_{\pm}(\epsilon)=\frac{1}{1+e^{\mp2\epsilon}}\,,
\end{equation}
where $\epsilon = Fl/2k_{B}T$ is the (scaled) external field intensity,
$k_B$ is Boltzmann's constant and $T$ is the temperature. The time
duration of these FPP jumps is~\cite{Slater93}:
\begin{equation}
    \label{e:tau1D}
    \tau (\epsilon)=\frac{\tanh\epsilon}{\epsilon}\tau_{B}\,,
\end{equation}
where $\tau_{B}$, the time duration $\tau (0)$ of a jump when no
external field is applied, is called the Brownian time.

Although Eqs.\eq{p1D} and\eq{tau1D} can be used to simulate
one-dimensional drift problems (the net velocity is then correct), they
erroneously generate a field-dependent diffusion coefficient for a free
particle, which is wrong. This failure is due to the lack of temporal
fluctuations in such a LMC algorithm (at each step, the particle would
jump either forward ($p_+$) or backward ($p_-$), and all jumps would
take the same time $\tau$). As mentioned above, it is possible to fix
this problem~\cite{PRL1} with a stochastic time step like $\tau\pm\Delta\tau$
where $\Delta \tau$ can also be calculated exactly within the framework
of FPP's~\cite{Slater93}: 
\begin{equation}
    \label{e:dtau}
     \Delta \tau(\epsilon) = \sqrt{\frac{\tanh\epsilon -
    \epsilon\sech^{2}\epsilon}{\epsilon^{3}}}\times\tau_{B}\,.
\end{equation}
However, the resulting algorithm can only be
used in Monte Carlo computer simulations because exact resolution
methods~\cite{Ogs9,EPJE1} require a common time step for all jumps.

Alternatively, temporal fluctuations can be introduced using a
probability $s^{\prime}$ to remain on the same lattice site during the
duration of a fixed time step $\tau^{\prime}$~\cite{PRL1}. Not moving
has for effect to create a dispersion of the time elapsed between two
actual jumps. In order to obtain the right free-solution diffusion
coefficient, we must have~\cite{PRL1} :
\begin{equation}
    \label{e:ss}
    s^{\prime} (\epsilon) = \frac{\coth\epsilon}{\epsilon}-\csch^{2}\epsilon\,.
\end{equation}
This modification also forces us to renormalize the other elements of
the LMC algorithm:
\begin{equation}
    \label{e:pp} p_{\pm}^{\prime}=(1-s^{\prime})p_{\pm}\,,
\end{equation}
\begin{equation}
    \label{e:tt} \tau^{\prime}=(1-s^{\prime})\tau\,.
\end{equation}
Equations\eq{ss} to\eq{tt} define a LMC algorithm that can be used
with Monte Carlo simulations (or exact numerical methods) to study
one-dimensional drift and diffusion problems. One can easily verify~\cite{PRL1} 
that it leads to the proper free-solution velocity
($v_{0}=\langle x \rangle / \tau^{\prime}=\epsilon l / \tau_{B}$) and
diffusion coefficient ($D_{0}=\langle \Delta x^{2}\rangle
/2\tau^{\prime}=l^{2}/2\tau_{B}$), while satisfying the Nernst-Einstein
relation $D_0 / v_0 = l/ \epsilon$. These equations will thus be the
starting point of our new multidimensional LMC algorithm.

\section{Extension to higher dimensions}\label{s:2D}
In principle, we can build a simple model for $d>1$ dimensions using
the elements of a one-dimensional biased random walk for the field axis
and those of an unbiased random-walk for each of the $d-1$ transverse
axes. Indeed, it is possible to fully decouple the motion along the
different spatial directions if the field is along a Cartesian axis.
Such an algorithm is divided into three steps:
\begin{enumerate}
\item First, we must select the jump axis, keeping in mind that the
particle should share its walking time equally between the $d$ spatial
directions. The probability to choose a given axis should thus be
inversely proportional to the mean time duration of a jump in this
direction (note that the time duration of a jump is shorter in the
field direction).
\item Secondly, the direction ($\pm$) of the jump must be selected.
\item Finally, the time duration of the jump must be computed and the
Monte Carlo clock must be advanced.
\end{enumerate}
There are several ways to implement these steps. The easiest way is to
use Eqs.\eq{p1D} to\eq{dtau}; in this case, the LMC clock must
advance by a stochastic increment $\tau \pm \Delta \tau$ each time a
jump is made along the field axis (in order to obtain the proper
temporal fluctuations, the clock does not advance otherwise). A
slightly more complicated way would be to use Eqs.\eq{ss} to\eq{tt};
again, the clock advances only when the jump is along the field axis,
but this choice has the advantage of not needing a stochastic time
increment. Although both of these implementations can easily be used
with computer simulations, they would not function with exact numerical
methods because of the way the clock is handled.

For exact numerical methods, an algorithm with a common time step and a
common clock for all spatial directions is required. We showed that it
is indeed possible to do this if we renormalize Eqs.\eq{p1D} and\eq{tau1D} properly~\cite{Ogs9}; this approach works for any dimension
$d>1$, but it can only be used to compute the exact velocity of the
particle since it neglects the temporal fluctuations. In order to also
include these fluctuations, one must start from Eqs.\eq{ss} to\eq{tt}
instead. Unfortunately, this can be done only in two dimensions since
the renormalization process gives negative probabilities when $d>2$~\cite{PRL1}.

Clearly, in order to derive a multi-dimensional LMC algorithm with a
fixed time-step, a common clock and the proper temporal fluctuations,
we need a major change to the basic assumptions of the LMC methodology.
In the next section, we propose to allow simultaneous jumps in all
spatial directions. This is a natural choice since LMC methods do
indeed assume that the motion of the particle is made of $d$ entirely
decoupled random-walks. Current LMC methods assume this decoupling to
be valid, but force the jumps to be sequential and not simultaneous.

\section{The need for a new set of moves}\label{s:new}
In our multi-dimensional algorithm~\cite{PRL1}, the LMC moves were the
standard unit jumps along one of the Cartesian axes, and a probability
to stay put was used to generate temporal fluctuations. Since moving
along a given axis actually contributes to temporal fluctuations along
all the other axes~\cite{PRL1}, the method fails for $d>2$ because the
transverse axes then provide an excess of temporal fluctuations. This
strongly suggests that the traditional sequential LMC moves are the
culprit. Sequential LMC moves are used solely for the sake of
simplicity, but they are a poor representation of the fact that real
particles move in all spatial directions at the same time. This
weakness is insignificant for unbiased diffusion, but it becomes a
roadblock in the presence of strong driving fields.

In order to resolve this problem, we thus suggest to employ a set of
moves that respect the simultaneous nature of the dynamics along each
of the $d$ axes. To generate a LMC algorithm for this new set of moves,
we will use our exact solution of the one-dimensional problem for each
of the $d$ directions.

\section{New $d$-dimensional LMC moves: the free-solution case}\label{s:free}
Our new LMC moves will include one jump attempt along each of $d$
spatial directions. The list will thus consist of $d\,! \times 3^d$
different moves since we must allow for all possible permutations of
the three fundamental jumps (of length $\pm l$ and $0$) used by the
exact one-dimensional model that we will be using for each axis. Note that the
external field must be parallel to one the Cartesian axes (we choose
the $x$-axis here). The dynamics is governed by $p_{\pm}^{\prime}$,
$s^{\prime}$ and $\tau^{\prime}$ in the $x$-direction (Eqs.\eq{ss} to\eq{tt}), whereas we can in principle use $p_{\bot}= \frac{1}{2}$ and
$\tau_{B}$ for the transverse directions because there is no need to
model the temporal fluctuations when there is no net drift in the given direction~\cite{PRL1}.

The optimal time step for our new moves is $\tau^{\prime}(\epsilon)$,
the duration of the fastest unit process. We thus have to rescale the
transverse probability $p_{\bot}$ accordingly:
\begin{equation}
    \label{e:rr}
    p_{\bot}^{\prime}=p_{\bot}\frac{\tau^{\prime}}{\tau_{B}}\,.
\end{equation}
This generates an arbitrary probability to stay put in the transverse
directions:
\begin{equation}
    \label{e:sy}
    s_{\bot}^{\prime} = 1-2p_{\bot}^{\prime} \,.
\end{equation}
In the zero-field limit, this probability gives:
\begin{equation}
    \left.s_{\bot}^{\prime}\right|_{\epsilon \rightarrow 0}
    = \frac{2}{3}
    = \left.s^{\prime}\right|_{\epsilon \rightarrow 0} \,.
\end{equation}
Therefore, the probability to stay put is the same in all the
directions in this limit, as it should. In the opposite limit $\epsilon
\rightarrow \infty$, we have:
\begin{equation}
    \left.s_{\bot}^{\prime}\right|_{\epsilon \rightarrow \infty} =
    1\,,
\end{equation}
and the jumps in the transverse directions become extremely rare, as
expected. Equations\eq{ss} to\eq{sy} are sufficient to build the
table of multi-dimensional moves and their different probabilities
since the $d$ directions are independent.

\begin{figure}[ht]
    \begin{center}
    \includegraphics[width=4.3in]{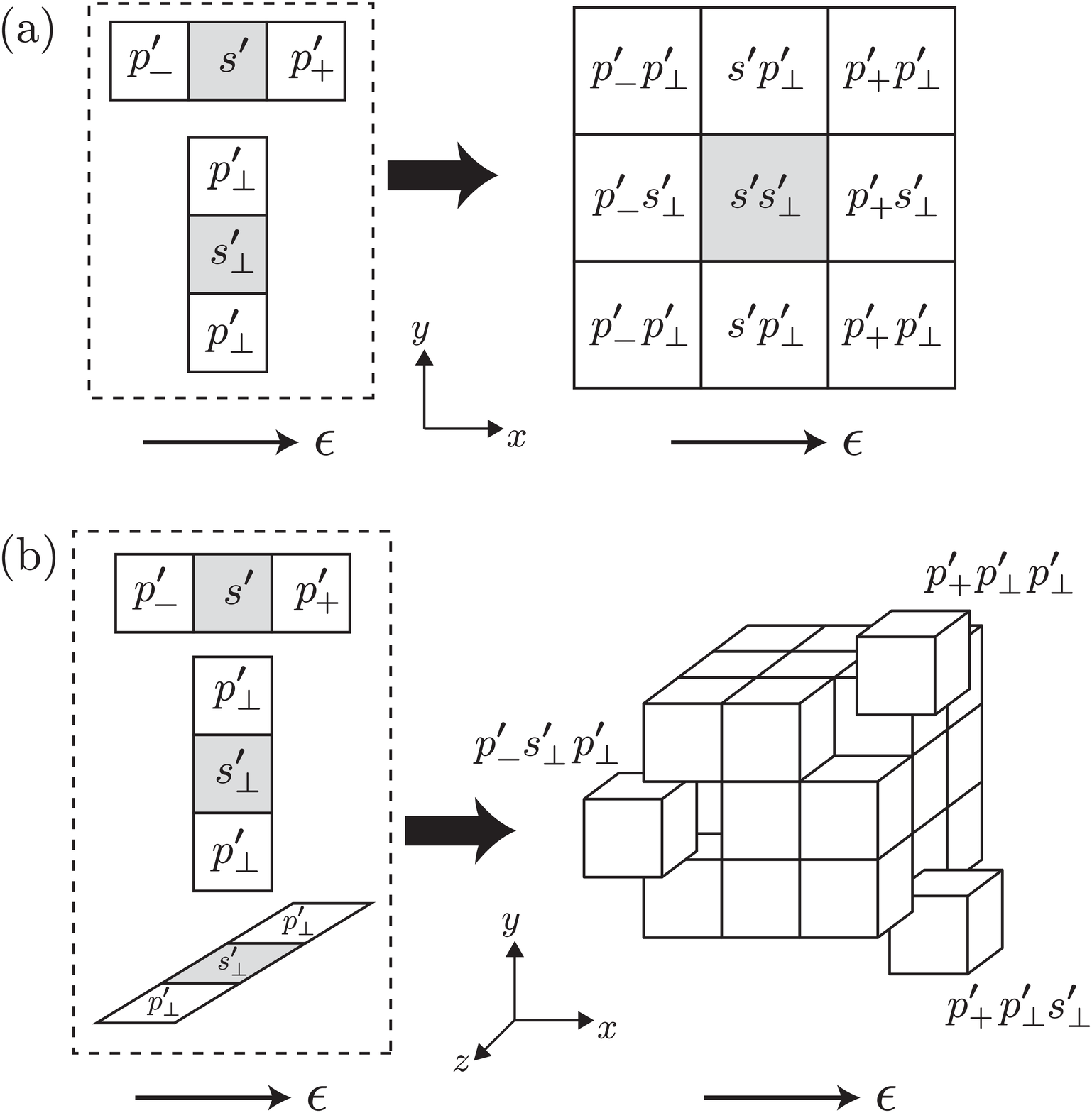}
        \caption{(a) Our new set of probabilities in two dimensions for an obstacle-free case (right) 
	is the result of the combination of two simultaneous one-dimensional processes (left).
        The grey site represents the position of the random walker before the transition.
        (b) Same as in (a) for a three-dimensional system (for clarity, we present only
        three of the final transition probabilities). }
        \label{f:schema}
    \end{center}
\end{figure}

Figure\fig{schema} illustrates the new LMC moves for the $2D$ and $3D$
cases in the absence of obstacles. The moves, all of duration
$\tau^{\prime}$, combine $d$ simultaneous one-dimensional processes and
include net displacements along lattice diagonals. The $d=2$ paths are
further defined in Table\tab{free}a; such a description of
the trajectories will be essential later to determine the dynamics in
the presence of obstacles. It is straightforward to extend this
approach to higher dimensions ($d>2$).

We can easily verify that this new set of LMC moves gives the right
free-solution velocity and diffusion coefficients for all dimensions $d
\geq 2$. If the field is pointing along the $x$-axis, the average
displacement per time step is ${\langle X\rangle}^{\prime}=
(p_{+}^{\prime}-p_{-}^{\prime})l$, while the average square
displacement is ${\langle X^{2}\rangle}^{\prime} =
(p_{+}^{\prime}+p_{-}^{\prime})l^{2}$. Using these results, we can
compute the free-solution velocity $v_{0_{x}}$ and diffusion
coefficient $D_{0_{x}}$:
\begin{equation}
    v_{0_{x}}
    = \frac{{\langle  X\rangle}^{\prime}}{\tau^{\prime}}
    = \frac{l\epsilon}{\tau_{B}}\,,
\end{equation}
and
\begin{equation}
    D_{0_{x}} = \frac{{\langle \Delta X^{2}\rangle}^{\prime}}{2\tau^{\prime}}
    = \frac{{\langle X^{2}\rangle}^{\prime}-{{\langle  X\rangle}^{\prime}}^{2}}{2\tau^{\prime}}
    = \frac{l^{2}}{2\tau_{B}}\,.
\end{equation}
One can also verify that $v_{0_{\bot}}=0$ and $D_{0_{\bot}} =
l^{2}/2\tau_{B}$. These are precisely the results that we expect.

Therefore, the model introduced here does work for all values of the
external field $\epsilon$ and all dimensions $d\geq 2$ in the absence of obstacles. The problems
faced in Ref.~\cite{PRL1} have been resolved by making the $d$
directions truly independent from each other and choosing
$\tau^{\prime}$ as the fundamental time step of the new LMC moves.

\section{New $d$-dimensional LMC moves: collisions with obstacles}\label{s:obs}
Since this new model works fine in free-solution, the next step is to
define how to deal with the presence of obstacles. The rule that we
follow in those cases where a move leads to a collision with an
obstacle is the same as before, i.e., such a jump is rejected and the
particle remains on the same site. In our algorithm, though, this means
that one (or more) of the $d$ sub-components of a $d$-dimensional move is
rejected. Therefore, the list of transition probabilities must take
into account all of the possible paths that the particle can follow
given the local geometry. A two-dimensional example is illustrated in
Table\tab{free}b. We see that the two different
trajectories that previously lead to the upper right corner (site $c$)
now lead to different final positions due to the rejection of one of
the two unit jumps that are involved. The final transition
probabilities for this particular case are listed in Table\tab{free}b.
Of course, all local distributions of obstacles can be studied using
the same systematic approach.

\section{New $d$-dimensional LMC moves: the continuum limit}\label{s:continuum}
In order to test our new set of LMC moves for systems with obstacles, we will compare its
predictions to those of our previous two-dimensional algorithm~\cite{PRL1} since we know that
both can properly reproduce the velocity and the diffusion coefficient
of a particle in the case of an obstacle-free system. However, the
different moves used by these two algorithms means that a true
comparison can only be made in the limit of the continuum since the
choice of moves always affects the result of a coarse-grained approach if there are obstacles.

The exact numerical method that we developed in collaboration with
Dorfman~\cite{EPJE1} is not limited to the previous set of LMC moves.
It can easily be modified to include other LMC moves, including
diagonal moves. Combining Dorfman's method~\cite{EPJE1,Dorfman1,Dorfman2} 
and our new LMC moves, we now have a way to compute the exact velocity and 
the exact diffusion coefficient of a particle in the presence of arbitrary 
driving field for any dimension $d \geq 2$.

\begin{figure}[ht]
    \begin{center}
    \includegraphics[width=5.5in]{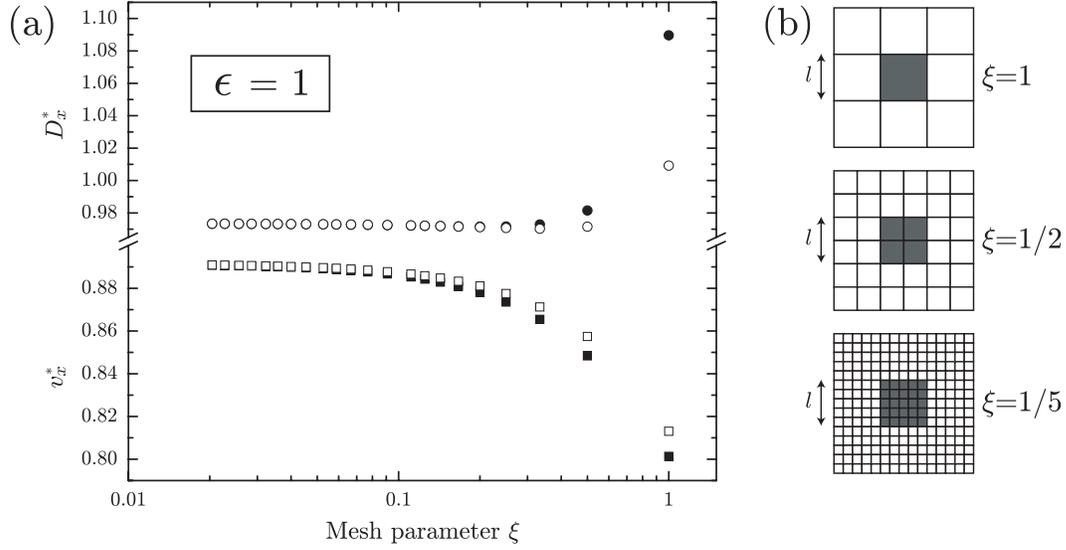}
        \caption{(a) Scaled velocity $v_{x}^{*}$ (squares) and diffusion coefficient
        $D_{x}^{*}$ (circles) vs the mesh size $\xi$ for $\epsilon=1$. These
        calculations were done using the algorithm presented in Ref.~\cite{EPJE1} (filled symbols)
        and the one proposed in this paper (empty symbols).
	(b) The obstacle is of size $l \times l$, the lattice is of size $3l\times 3l$ (with periodic 
	boundary conditions), and the particle (not shown) is of size $\xi l \times \xi l$. The system 
	is shown for three different values of the mesh size parameter $\xi$. }
        \label{f:continuum}
    \end{center}
\end{figure}

We thus studied the system shown in Fig.\fig{continuum}b
using both algorithms, and we repeated the calculation for different
lattice parameters $\xi l$ (with $0<\xi\leq 1$) while the obstacle size
($l$) remained constant (the surface concentration of obstacles is thus
kept constant at $C=1/9$). The limit of the continuum corresponds to
$\xi\rightarrow 0$. We compared the velocities and diffusion
coefficients along the field-axis obtained with both algorithms over a
wide range of $\xi$. Note that the value of the external scaled field
$\epsilon$, which is proportional to the lattice parameter ($\epsilon =
Fl/2k_{B}T$), has to be rescaled by the factor $\xi$. Figure\fig{continuum}a 
presents the data for both algorithms for a nominal
field intensity $\epsilon=1$. We clearly see that the two approaches
converge perfectly in the $\xi \rightarrow 0$ limit. Interestingly, the
new algorithm converges slightly faster towards the asymptotic
continuum value. This is explained by the fact that the diagonal transitions reduce the 
number of successive collisions made by a random-walker when it is trapped behind an 
obstacle at high field.

\section{Conclusion} \label{s:dis}
Conventional three-dimensional LMC algorithms cannot be used to study
both the mean velocity and the diffusion coefficient of a Brownian
particle if the time step has to be constant (as required by exact
numerical methods). This limitation is due to the fact that these
algorithms only allow jumps to be made along one axis at each time
step. Such unit jumps make it impossible to obtain the proper temporal
fluctuations that are key to getting the right diffusion coefficient.

We propose that LMC moves should actually respect the fact that all of
the $d$ spatial dimensions are fully independent. This means that each
move should include a component along each of these dimensions. This
complete dimensional decoupling allows us to conserve the proper
temporal fluctuations and hence to reproduce the correct diffusion
process even in the presence of an external field of arbitrary
amplitude. This approach leads to a slightly more complicated analysis
of particle-obstacle collisions, but this is still compatible with the
exact numerical methods developed elsewhere~\cite{PRL1}.

The new LMC algorithm presented in this paper opens the door to
numerous coarse-grained simulation and numerical studies that were not
possible before because previous algorithms were restricted to low
field intensities.

\section*{Acknowledgments}
This work was supported by a Discovery Grant from the Natural Science
and Engineering Research Council (\emph{NSERC}) of Canada to GWS. MGG
was supported by a \emph{NSERC} scholarship, an excellence scholarship 
from the University of Ottawa and a Strategic Areas of Development (SAD) 
scholarship from the University of Ottawa. 


\newpage

\newpage

\newpage
\begin{table}[ht]
  \begin{scriptsize}
  \centering   
  \caption{Listing of all the possible trajectories and their transition probabilities in two dimensions for the free-solution case (a) and an example of obstacle obstruction (b). }\label{t:free}
      \setlength{\tabcolsep}{2.2mm}
    \begin{tabular}{ccc}
    \\
    \begin{normalsize} \hspace{-8mm}(a) Free-solution case \end{normalsize}& &\begin{normalsize} \hspace{-8mm} (b) Obstacle case \end{normalsize}\\
    \begin{tabular}{cccc}\hline\hline
        \rule[-6mm]{0mm}{14mm}
        \begin{minipage}{0.7cm}  \renewcommand{\baselinestretch}{0.0}\scriptsize \begin{center} $1^{st}$ \\ jump \end{center}  \end{minipage} & \begin{minipage}{0.7cm}  \renewcommand{\baselinestretch}{0.0}\scriptsize \begin{center} $2^{nd}$ \\ jump \end{center}  \end{minipage}  & \begin{minipage}{1.0cm}  \renewcommand{\baselinestretch}{0.05}\scriptsize \begin{center} final \\ position \end{center}  \end{minipage} & \begin{minipage}{1.8cm} \renewcommand{\baselinestretch}{0.85}\scriptsize \begin{center} transition \\ probability \end{center}  \end{minipage} \\ \hline\hline
        \setlength{\tabcolsep}{0mm} \begin{tabular}{c} $p_{-}^{\prime}$ \\ $p_{\bot}^{\prime}$ \end{tabular} & \begin{tabular}{c} $p_{\bot}^{\prime}$  \\ $p_{-}^{\prime}$ \end{tabular} & $a$ & $p_{-}^{\prime}p_{\bot}^{\prime}$ \\ \hline
        \setlength{\tabcolsep}{0mm} \begin{tabular}{c} $s^{\prime}$     \\ $p_{\bot}^{\prime}$ \end{tabular} & \begin{tabular}{c} $p_{\bot}^{\prime}$  \\ $s^{\prime}$ \end{tabular}     & $b$ & $s^{\prime}p_{\bot}^{\prime}$ \\ \hline
        \setlength{\tabcolsep}{0mm} \begin{tabular}{c} $p_{+}^{\prime}$ \\ $p_{\bot}^{\prime}$ \end{tabular} & \begin{tabular}{c} $p_{\bot}^{\prime}$  \\ $p_{+}^{\prime}$ \end{tabular} & $c$ & $p_{+}^{\prime}p_{\bot}^{\prime}$ \\ \hline
        \setlength{\tabcolsep}{0mm} \begin{tabular}{c} $p_{-}^{\prime}$ \\ $s_{\bot}^{\prime}$ \end{tabular} & \begin{tabular}{c} $s_{\bot}^{\prime}$  \\ $p_{-}^{\prime}$ \end{tabular} & $d$ & $p_{-}^{\prime}s_{\bot}^{\prime}$ \\ \hline
        \setlength{\tabcolsep}{0mm} \begin{tabular}{c} $s^{\prime}$     \\ $s_{\bot}^{\prime}$ \end{tabular} & \begin{tabular}{c} $s_{\bot}^{\prime}$  \\ $s^{\prime}$ \end{tabular}     & $e$ & $s^{\prime}s_{\bot}^{\prime}$ \\ \hline
        \setlength{\tabcolsep}{0mm} \begin{tabular}{c} $p_{+}^{\prime}$ \\ $s_{\bot}^{\prime}$ \end{tabular} & \begin{tabular}{c} $s_{\bot}^{\prime}$  \\ $p_{+}^{\prime}$ \end{tabular} & $f$ & $p_{+}^{\prime}s_{\bot}^{\prime}$ \\ \hline
        \setlength{\tabcolsep}{0mm} \begin{tabular}{c} $p_{-}^{\prime}$ \\ $p_{\bot}^{\prime}$ \end{tabular} & \begin{tabular}{c} $p_{\bot}^{\prime}$  \\ $p_{-}^{\prime}$ \end{tabular} & $g$ & $p_{-}^{\prime}p_{\bot}^{\prime}$ \\ \hline
        \setlength{\tabcolsep}{0mm} \begin{tabular}{c} $s^{\prime}$     \\ $p_{\bot}^{\prime}$ \end{tabular} & \begin{tabular}{c} $p_{\bot}^{\prime}$  \\ $s^{\prime}$ \end{tabular}     & $h$ & $s^{\prime}p_{\bot}^{\prime}$ \\ \hline
        \setlength{\tabcolsep}{0mm} \begin{tabular}{c} $p_{+}^{\prime}$ \\ $p_{\bot}^{\prime}$ \end{tabular} & \begin{tabular}{c} $p_{\bot}^{\prime}$  \\ $p_{+}^{\prime}$ \end{tabular} & $i$ & $p_{+}^{\prime}p_{\bot}^{\prime}$ \\ \hline
    	\hline
   
	\multicolumn{4}{c}{ \includegraphics[width=1.25in]{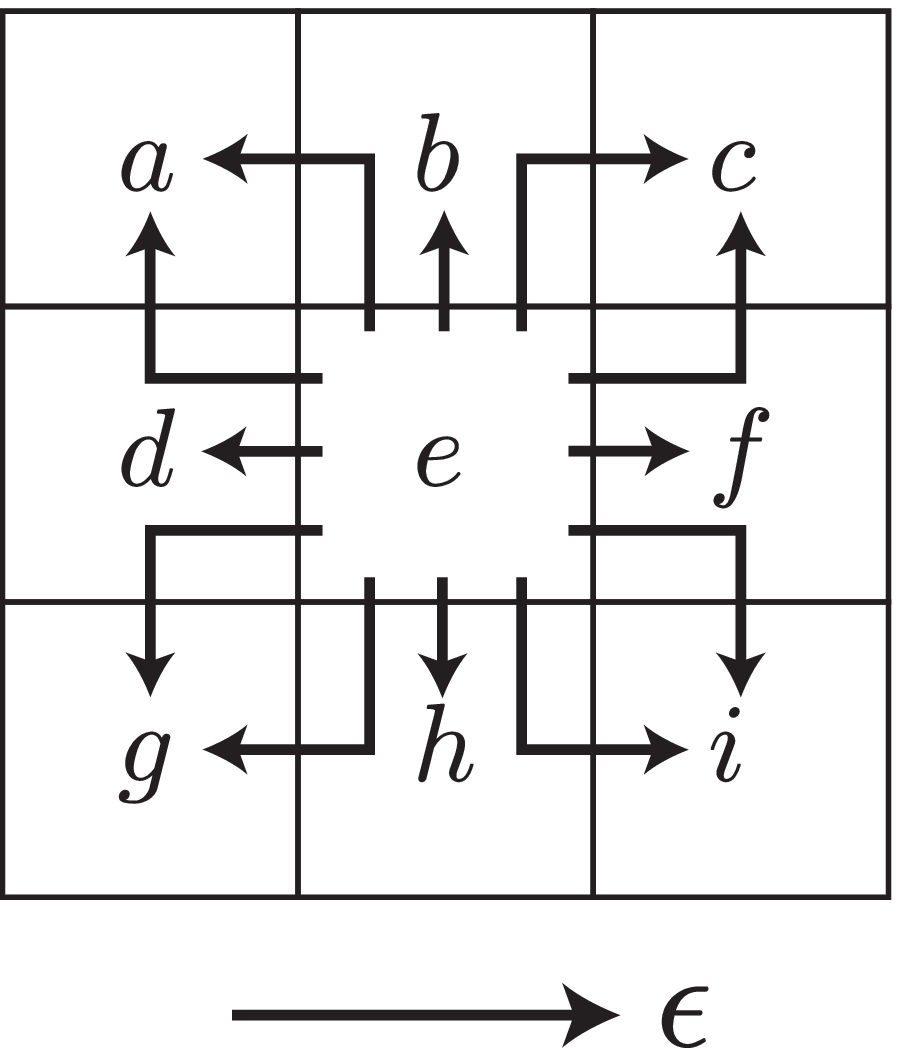} } \\ \hline\hline
    \end{tabular}
    & & 
    \begin{tabular}{cccc}\hline\hline
        \rule[-6mm]{0mm}{14mm}
        \begin{minipage}{0.7cm}  \renewcommand{\baselinestretch}{0.0}\scriptsize \begin{center} $1^{st}$ \\ jump \end{center}  \end{minipage} & \begin{minipage}{0.7cm}  \renewcommand{\baselinestretch}{0.0}\scriptsize \begin{center} $2^{nd}$ \\ jump \end{center}  \end{minipage}  & \begin{minipage}{1.0cm}  \renewcommand{\baselinestretch}{0.05}\scriptsize \begin{center} final \\ position \end{center}  \end{minipage} & \begin{minipage}{1.8cm} \renewcommand{\baselinestretch}{0.85}\scriptsize \begin{center} transition \\ probability \end{center}  \end{minipage} \\ \hline\hline
        \setlength{\tabcolsep}{0mm} \begin{tabular}{c} $p_{-}^{\prime}$ \\ $p_{\bot}^{\prime}$ \end{tabular} & \begin{tabular}{c} $p_{\bot}^{\prime}$  \\ $p_{-}^{\prime}$ \end{tabular} & $a$ & $p_{-}^{\prime}p_{\bot}^{\prime}$ \\ \hline
        \setlength{\tabcolsep}{0mm} \begin{tabular}{c} $s^{\prime}$     \\ $p_{\bot}^{\prime}$ \\ $p_{\bot}^{\prime}$ \end{tabular} & \begin{tabular}{c} $p_{\bot}^{\prime}$  \\ $s^{\prime}$ \\ $p_{+}^{\prime}$ \end{tabular}     & $b$ & $\displaystyle s^{\prime}p_{\bot}^{\prime}+\frac{p_{+}^{\prime}p_{\bot}^{\prime}}{2}$ \\ \hline
        \setlength{\tabcolsep}{0mm} \begin{tabular}{c} $p_{-}^{\prime}$ \\ $s_{\bot}^{\prime}$ \end{tabular} & \begin{tabular}{c} $s_{\bot}^{\prime}$  \\ $p_{-}^{\prime}$ \end{tabular} & $d$ & $p_{-}^{\prime}s_{\bot}^{\prime}$ \\ \hline
        \setlength{\tabcolsep}{0mm} \begin{tabular}{c} $s^{\prime}$     \\ $s_{\bot}^{\prime}$ \end{tabular} & \begin{tabular}{c} $s_{\bot}^{\prime}$  \\ $s^{\prime}$ \end{tabular}     & $e$ & $s^{\prime}s_{\bot}^{\prime}$ \\ \hline
        \setlength{\tabcolsep}{0mm} \begin{tabular}{c} $p_{+}^{\prime}$ \\ $s_{\bot}^{\prime}$ \\ $p_{+}^{\prime}$ \end{tabular} & \begin{tabular}{c} $s_{\bot}^{\prime}$  \\ $p_{+}^{\prime}$ \\ $p_{\bot}^{\prime}$\end{tabular} & $f$ & $\displaystyle p_{+}^{\prime}s_{\bot}^{\prime}+\frac{p_{+}^{\prime}p_{\bot}^{\prime}}{2}$ \\ \hline
        \setlength{\tabcolsep}{0mm} \begin{tabular}{c} $p_{-}^{\prime}$ \\ $p_{\bot}^{\prime}$ \end{tabular} & \begin{tabular}{c} $p_{\bot}^{\prime}$  \\ $p_{-}^{\prime}$ \end{tabular} & $g$ & $p_{-}^{\prime}p_{\bot}^{\prime}$ \\ \hline
        \setlength{\tabcolsep}{0mm} \begin{tabular}{c} $s^{\prime}$     \\ $p_{\bot}^{\prime}$ \end{tabular} & \begin{tabular}{c} $p_{\bot}^{\prime}$  \\ $s^{\prime}$ \end{tabular}     & $h$ & $s^{\prime}p_{\bot}^{\prime}$ \\ \hline
        \setlength{\tabcolsep}{0mm} \begin{tabular}{c} $p_{+}^{\prime}$ \\ $p_{\bot}^{\prime}$ \end{tabular} & \begin{tabular}{c} $p_{\bot}^{\prime}$  \\ $p_{+}^{\prime}$ \end{tabular} & $i$ & $p_{+}^{\prime}p_{\bot}^{\prime}$ \\ \hline
    	\hline
   
	\multicolumn{4}{c}{ \includegraphics[width=1.25in]{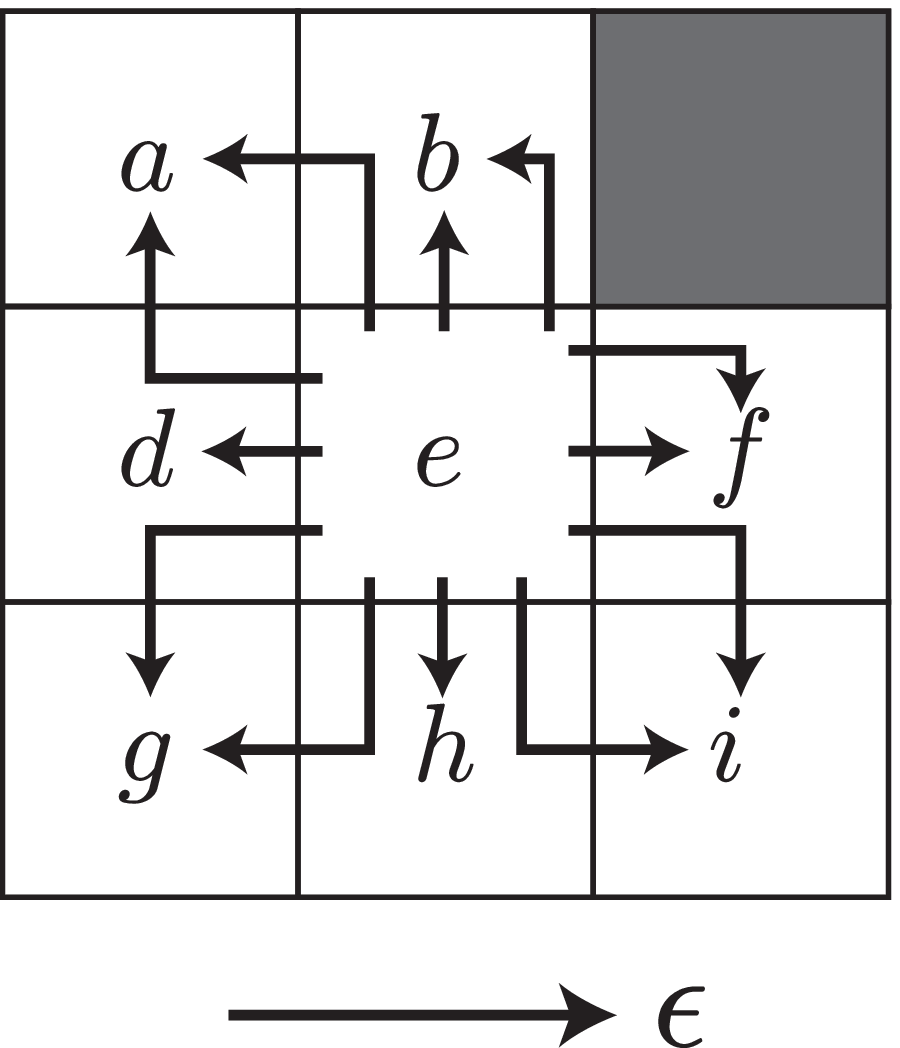} } \\ \hline\hline
    \end{tabular}
    \end{tabular}
    \end{scriptsize}
\end{table}

\end{document}